\documentclass[12pt,a4paper,superscriptaddress]{revtex4}
\usepackage{graphicx}
\usepackage{graphics}
\usepackage{amsmath}
\usepackage{dcolumn}
\usepackage{amssymb}
\usepackage{bm}
\usepackage[latin1]{inputenc}
\newcommand{\be}{\begin{equation}}
\newcommand{\ee}{\end{equation}}
\newcommand{\bea}{\begin{eqnarray}}
\newcommand{\eea}{\end{eqnarray}}

\begin{document}
\immediate\write16{<<WARNING: LINEDRAW macros work with emTeX-dvivers
                    and other drivers supporting emTeX \special's
                    (dviscr, dvihplj, dvidot, dvips, dviwin, etc.) >>}

\newdimen\Lengthunit       \Lengthunit  = 1.5cm
\newcount\Nhalfperiods     \Nhalfperiods= 9
\newcount\magnitude        \magnitude = 1000

\catcode`\*=11
\newdimen\L*   \newdimen\d*   \newdimen\d**
\newdimen\dm*  \newdimen\dd*  \newdimen\dt*
\newdimen\a*   \newdimen\b*   \newdimen\c*
\newdimen\a**  \newdimen\b**
\newdimen\xL*  \newdimen\yL*
\newdimen\rx*  \newdimen\ry*
\newdimen\tmp* \newdimen\linwid*

\newcount\k*   \newcount\l*   \newcount\m*
\newcount\k**  \newcount\l**  \newcount\m**
\newcount\n*   \newcount\dn*  \newcount\r*
\newcount\N*   \newcount\*one \newcount\*two  \*one=1 \*two=2
\newcount\*ths \*ths=1000
\newcount\angle*  \newcount\q*  \newcount\q**
\newcount\angle** \angle**=0
\newcount\sc*     \sc*=0

\newtoks\cos*  \cos*={1}
\newtoks\sin*  \sin*={0}

\catcode`\[=13

\def\rotate(#1){\advance\angle**#1\angle*=\angle**
\q**=\angle*\ifnum\q**<0\q**=-\q**\fi
\ifnum\q**>360\q*=\angle*\divide\q*360\multiply\q*360\advance\angle*-\q*\fi
\ifnum\angle*<0\advance\angle*360\fi\q**=\angle*\divide\q**90\q**=\q**
\def\sgcos*{+}\def\sgsin*{+}\relax
\ifcase\q**\or
 \def\sgcos*{-}\def\sgsin*{+}\or
 \def\sgcos*{-}\def\sgsin*{-}\or
 \def\sgcos*{+}\def\sgsin*{-}\else\fi
\q*=\q**
\multiply\q*90\advance\angle*-\q*
\ifnum\angle*>45\sc*=1\angle*=-\angle*\advance\angle*90\else\sc*=0\fi
\def[##1,##2]{\ifnum\sc*=0\relax
\edef\cs*{\sgcos*.##1}\edef\sn*{\sgsin*.##2}\ifcase\q**\or
 \edef\cs*{\sgcos*.##2}\edef\sn*{\sgsin*.##1}\or
 \edef\cs*{\sgcos*.##1}\edef\sn*{\sgsin*.##2}\or
 \edef\cs*{\sgcos*.##2}\edef\sn*{\sgsin*.##1}\else\fi\else
\edef\cs*{\sgcos*.##2}\edef\sn*{\sgsin*.##1}\ifcase\q**\or
 \edef\cs*{\sgcos*.##1}\edef\sn*{\sgsin*.##2}\or
 \edef\cs*{\sgcos*.##2}\edef\sn*{\sgsin*.##1}\or
 \edef\cs*{\sgcos*.##1}\edef\sn*{\sgsin*.##2}\else\fi\fi
\cos*={\cs*}\sin*={\sn*}\global\edef\gcos*{\cs*}\global\edef\gsin*{\sn*}}\relax
\ifcase\angle*[9999,0]\or
[999,017]\or[999,034]\or[998,052]\or[997,069]\or[996,087]\or
[994,104]\or[992,121]\or[990,139]\or[987,156]\or[984,173]\or
[981,190]\or[978,207]\or[974,224]\or[970,241]\or[965,258]\or
[961,275]\or[956,292]\or[951,309]\or[945,325]\or[939,342]\or
[933,358]\or[927,374]\or[920,390]\or[913,406]\or[906,422]\or
[898,438]\or[891,453]\or[882,469]\or[874,484]\or[866,499]\or
[857,515]\or[848,529]\or[838,544]\or[829,559]\or[819,573]\or
[809,587]\or[798,601]\or[788,615]\or[777,629]\or[766,642]\or
[754,656]\or[743,669]\or[731,681]\or[719,694]\or[707,707]\or
\else[9999,0]\fi}

\catcode`\[=12

\def\GRAPH(hsize=#1)#2{\hbox to #1\Lengthunit{#2\hss}}

\def\Linewidth#1{\global\linwid*=#1\relax
\global\divide\linwid*10\global\multiply\linwid*\mag
\global\divide\linwid*100\special{em:linewidth \the\linwid*}}

\Linewidth{.4pt}
\def\sm*{\special{em:moveto}}
\def\sl*{\special{em:lineto}}
\let\moveto=\sm*
\let\lineto=\sl*
\newbox\spm*   \newbox\spl*
\setbox\spm*\hbox{\sm*}
\setbox\spl*\hbox{\sl*}

\def\mov#1(#2,#3)#4{\rlap{\L*=#1\Lengthunit
\xL*=#2\L* \yL*=#3\L*
\xL*=\xscale\xL* \yL*=\yscale\yL*
\rx* \the\cos*\xL* \tmp* \the\sin*\yL* \advance\rx*-\tmp*
\ry* \the\cos*\yL* \tmp* \the\sin*\xL* \advance\ry*\tmp*
\kern\rx*\raise\ry*\hbox{#4}}}

\def\rmov*(#1,#2)#3{\rlap{\xL*=#1\yL*=#2\relax
\rx* \the\cos*\xL* \tmp* \the\sin*\yL* \advance\rx*-\tmp*
\ry* \the\cos*\yL* \tmp* \the\sin*\xL* \advance\ry*\tmp*
\kern\rx*\raise\ry*\hbox{#3}}}

\def\lin#1(#2,#3){\rlap{\sm*\mov#1(#2,#3){\sl*}}}

\def\arr*(#1,#2,#3){\rmov*(#1\dd*,#1\dt*){\sm*
\rmov*(#2\dd*,#2\dt*){\rmov*(#3\dt*,-#3\dd*){\sl*}}\sm*
\rmov*(#2\dd*,#2\dt*){\rmov*(-#3\dt*,#3\dd*){\sl*}}}}

\def\arrow#1(#2,#3){\rlap{\lin#1(#2,#3)\mov#1(#2,#3){\relax
\d**=-.012\Lengthunit\dd*=#2\d**\dt*=#3\d**
\arr*(1,10,4)\arr*(3,8,4)\arr*(4.8,4.2,3)}}}

\def\arrlin#1(#2,#3){\rlap{\L*=#1\Lengthunit\L*=.5\L*
\lin#1(#2,#3)\rmov*(#2\L*,#3\L*){\arrow.1(#2,#3)}}}

\def\dasharrow#1(#2,#3){\rlap{{\Lengthunit=0.9\Lengthunit
\dashlin#1(#2,#3)\mov#1(#2,#3){\sm*}}\mov#1(#2,#3){\sl*
\d**=-.012\Lengthunit\dd*=#2\d**\dt*=#3\d**
\arr*(1,10,4)\arr*(3,8,4)\arr*(4.8,4.2,3)}}}

\def\clap#1{\hbox to 0pt{\hss #1\hss}}

\def\ind(#1,#2)#3{\rlap{\L*=.1\Lengthunit
\xL*=#1\L* \yL*=#2\L*
\rx* \the\cos*\xL* \tmp* \the\sin*\yL* \advance\rx*-\tmp*
\ry* \the\cos*\yL* \tmp* \the\sin*\xL* \advance\ry*\tmp*
\kern\rx*\raise\ry*\hbox{\lower2pt\clap{$#3$}}}}

\def\sh*(#1,#2)#3{\rlap{\dm*=\the\n*\d**
\xL*=\xscale\dm* \yL*=\yscale\dm* \xL*=#1\xL* \yL*=#2\yL*
\rx* \the\cos*\xL* \tmp* \the\sin*\yL* \advance\rx*-\tmp*
\ry* \the\cos*\yL* \tmp* \the\sin*\xL* \advance\ry*\tmp*
\kern\rx*\raise\ry*\hbox{#3}}}

\def\calcnum*#1(#2,#3){\a*=1000sp\b*=1000sp\a*=#2\a*\b*=#3\b*
\ifdim\a*<0pt\a*-\a*\fi\ifdim\b*<0pt\b*-\b*\fi
\ifdim\a*>\b*\c*=.96\a*\advance\c*.4\b*
\else\c*=.96\b*\advance\c*.4\a*\fi
\k*\a*\multiply\k*\k*\l*\b*\multiply\l*\l*
\m*\k*\advance\m*\l*\n*\c*\r*\n*\multiply\n*\n*
\dn*\m*\advance\dn*-\n*\divide\dn*2\divide\dn*\r*
\advance\r*\dn*
\c*=\the\Nhalfperiods5sp\c*=#1\c*\ifdim\c*<0pt\c*-\c*\fi
\multiply\c*\r*\N*\c*\divide\N*10000}

\def\dashlin#1(#2,#3){\rlap{\calcnum*#1(#2,#3)\relax
\d**=#1\Lengthunit\ifdim\d**<0pt\d**-\d**\fi
\divide\N*2\multiply\N*2\advance\N*\*one
\divide\d**\N*\sm*\n*\*one\sh*(#2,#3){\sl*}\loop
\advance\n*\*one\sh*(#2,#3){\sm*}\advance\n*\*one
\sh*(#2,#3){\sl*}\ifnum\n*<\N*\repeat}}

\def\dashdotlin#1(#2,#3){\rlap{\calcnum*#1(#2,#3)\relax
\d**=#1\Lengthunit\ifdim\d**<0pt\d**-\d**\fi
\divide\N*2\multiply\N*2\advance\N*1\multiply\N*2\relax
\divide\d**\N*\sm*\n*\*two\sh*(#2,#3){\sl*}\loop
\advance\n*\*one\sh*(#2,#3){\kern-1.48pt\lower.5pt\hbox{\rm.}}\relax
\advance\n*\*one\sh*(#2,#3){\sm*}\advance\n*\*two
\sh*(#2,#3){\sl*}\ifnum\n*<\N*\repeat}}

\def\shl*(#1,#2)#3{\kern#1#3\lower#2#3\hbox{\unhcopy\spl*}}

\def\trianglin#1(#2,#3){\rlap{\toks0={#2}\toks1={#3}\calcnum*#1(#2,#3)\relax
\dd*=.57\Lengthunit\dd*=#1\dd*\divide\dd*\N*
\divide\dd*\*ths \multiply\dd*\magnitude
\d**=#1\Lengthunit\ifdim\d**<0pt\d**-\d**\fi
\multiply\N*2\divide\d**\N*\sm*\n*\*one\loop
\shl**{\dd*}\dd*-\dd*\advance\n*2\relax
\ifnum\n*<\N*\repeat\n*\N*\shl**{0pt}}}

\def\wavelin#1(#2,#3){\rlap{\toks0={#2}\toks1={#3}\calcnum*#1(#2,#3)\relax
\dd*=.23\Lengthunit\dd*=#1\dd*\divide\dd*\N*
\divide\dd*\*ths \multiply\dd*\magnitude
\d**=#1\Lengthunit\ifdim\d**<0pt\d**-\d**\fi
\multiply\N*4\divide\d**\N*\sm*\n*\*one\loop
\shl**{\dd*}\dt*=1.3\dd*\advance\n*\*one
\shl**{\dt*}\advance\n*\*one
\shl**{\dd*}\advance\n*\*two
\dd*-\dd*\ifnum\n*<\N*\repeat\n*\N*\shl**{0pt}}}

\def\w*lin(#1,#2){\rlap{\toks0={#1}\toks1={#2}\d**=\Lengthunit\dd*=-.12\d**
\divide\dd*\*ths \multiply\dd*\magnitude
\N*8\divide\d**\N*\sm*\n*\*one\loop
\shl**{\dd*}\dt*=1.3\dd*\advance\n*\*one
\shl**{\dt*}\advance\n*\*one
\shl**{\dd*}\advance\n*\*one
\shl**{0pt}\dd*-\dd*\advance\n*1\ifnum\n*<\N*\repeat}}

\def\l*arc(#1,#2)[#3][#4]{\rlap{\toks0={#1}\toks1={#2}\d**=\Lengthunit
\dd*=#3.037\d**\dd*=#4\dd*\dt*=#3.049\d**\dt*=#4\dt*\ifdim\d**>10mm\relax
\d**=.25\d**\n*\*one\shl**{-\dd*}\n*\*two\shl**{-\dt*}\n*3\relax
\shl**{-\dd*}\n*4\relax\shl**{0pt}\else
\ifdim\d**>5mm\d**=.5\d**\n*\*one\shl**{-\dt*}\n*\*two
\shl**{0pt}\else\n*\*one\shl**{0pt}\fi\fi}}

\def\d*arc(#1,#2)[#3][#4]{\rlap{\toks0={#1}\toks1={#2}\d**=\Lengthunit
\dd*=#3.037\d**\dd*=#4\dd*\d**=.25\d**\sm*\n*\*one\shl**{-\dd*}\relax
\n*3\relax\sh*(#1,#2){\xL*=\xscale\dd*\yL*=\yscale\dd*
\kern#2\xL*\lower#1\yL*\hbox{\sm*}}\n*4\relax\shl**{0pt}}}

\def\shl**#1{\c*=\the\n*\d**\d*=#1\relax
\a*=\the\toks0\c*\b*=\the\toks1\d*\advance\a*-\b*
\b*=\the\toks1\c*\d*=\the\toks0\d*\advance\b*\d*
\a*=\xscale\a*\b*=\yscale\b*
\rx* \the\cos*\a* \tmp* \the\sin*\b* \advance\rx*-\tmp*
\ry* \the\cos*\b* \tmp* \the\sin*\a* \advance\ry*\tmp*
\raise\ry*\rlap{\kern\rx*\unhcopy\spl*}}

\def\wlin*#1(#2,#3)[#4]{\rlap{\toks0={#2}\toks1={#3}\relax
\c*=#1\l*\c*\c*=.01\Lengthunit\m*\c*\divide\l*\m*
\c*=\the\Nhalfperiods5sp\multiply\c*\l*\N*\c*\divide\N*\*ths
\divide\N*2\multiply\N*2\advance\N*\*one
\dd*=.002\Lengthunit\dd*=#4\dd*\multiply\dd*\l*\divide\dd*\N*
\divide\dd*\*ths \multiply\dd*\magnitude
\d**=#1\multiply\N*4\divide\d**\N*\sm*\n*\*one\loop
\shl**{\dd*}\dt*=1.3\dd*\advance\n*\*one
\shl**{\dt*}\advance\n*\*one
\shl**{\dd*}\advance\n*\*two
\dd*-\dd*\ifnum\n*<\N*\repeat\n*\N*\shl**{0pt}}}

\def\wavebox#1{\setbox0\hbox{#1}\relax
\a*=\wd0\advance\a*14pt\b*=\ht0\advance\b*\dp0\advance\b*14pt\relax
\hbox{\kern9pt\relax
\rmov*(0pt,\ht0){\rmov*(-7pt,7pt){\wlin*\a*(1,0)[+]\wlin*\b*(0,-1)[-]}}\relax
\rmov*(\wd0,-\dp0){\rmov*(7pt,-7pt){\wlin*\a*(-1,0)[+]\wlin*\b*(0,1)[-]}}\relax
\box0\kern9pt}}

\def\rectangle#1(#2,#3){\relax
\lin#1(#2,0)\lin#1(0,#3)\mov#1(0,#3){\lin#1(#2,0)}\mov#1(#2,0){\lin#1(0,#3)}}

\def\dashrectangle#1(#2,#3){\dashlin#1(#2,0)\dashlin#1(0,#3)\relax
\mov#1(0,#3){\dashlin#1(#2,0)}\mov#1(#2,0){\dashlin#1(0,#3)}}

\def\waverectangle#1(#2,#3){\L*=#1\Lengthunit\a*=#2\L*\b*=#3\L*
\ifdim\a*<0pt\a*-\a*\def\x*{-1}\else\def\x*{1}\fi
\ifdim\b*<0pt\b*-\b*\def\y*{-1}\else\def\y*{1}\fi
\wlin*\a*(\x*,0)[-]\wlin*\b*(0,\y*)[+]\relax
\mov#1(0,#3){\wlin*\a*(\x*,0)[+]}\mov#1(#2,0){\wlin*\b*(0,\y*)[-]}}

\def\calcparab*{\ifnum\n*>\m*\k*\N*\advance\k*-\n*\else\k*\n*\fi
\a*=\the\k* sp\a*=10\a*\b*\dm*\advance\b*-\a*\k*\b*
\a*=\the\*ths\b*\divide\a*\l*\multiply\a*\k*
\divide\a*\l*\k*\*ths\r*\a*\advance\k*-\r*\dt*=\the\k*\L*}

\def\arcto#1(#2,#3)[#4]{\rlap{\toks0={#2}\toks1={#3}\calcnum*#1(#2,#3)\relax
\dm*=135sp\dm*=#1\dm*\d**=#1\Lengthunit\ifdim\dm*<0pt\dm*-\dm*\fi
\multiply\dm*\r*\a*=.3\dm*\a*=#4\a*\ifdim\a*<0pt\a*-\a*\fi
\advance\dm*\a*\N*\dm*\divide\N*10000\relax
\divide\N*2\multiply\N*2\advance\N*\*one
\L*=-.25\d**\L*=#4\L*\divide\d**\N*\divide\L*\*ths
\m*\N*\divide\m*2\dm*=\the\m*5sp\l*\dm*\sm*\n*\*one\loop
\calcparab*\shl**{-\dt*}\advance\n*1\ifnum\n*<\N*\repeat}}

\def\arrarcto#1(#2,#3)[#4]{\L*=#1\Lengthunit\L*=.54\L*
\arcto#1(#2,#3)[#4]\rmov*(#2\L*,#3\L*){\d*=.457\L*\d*=#4\d*\d**-\d*
\rmov*(#3\d**,#2\d*){\arrow.02(#2,#3)}}}

\def\dasharcto#1(#2,#3)[#4]{\rlap{\toks0={#2}\toks1={#3}\relax
\calcnum*#1(#2,#3)\dm*=\the\N*5sp\a*=.3\dm*\a*=#4\a*\ifdim\a*<0pt\a*-\a*\fi
\advance\dm*\a*\N*\dm*
\divide\N*20\multiply\N*2\advance\N*1\d**=#1\Lengthunit
\L*=-.25\d**\L*=#4\L*\divide\d**\N*\divide\L*\*ths
\m*\N*\divide\m*2\dm*=\the\m*5sp\l*\dm*
\sm*\n*\*one\loop\calcparab*
\shl**{-\dt*}\advance\n*1\ifnum\n*>\N*\else\calcparab*
\sh*(#2,#3){\xL*=#3\dt* \yL*=#2\dt*
\rx* \the\cos*\xL* \tmp* \the\sin*\yL* \advance\rx*\tmp*
\ry* \the\cos*\yL* \tmp* \the\sin*\xL* \advance\ry*-\tmp*
\kern\rx*\lower\ry*\hbox{\sm*}}\fi
\advance\n*1\ifnum\n*<\N*\repeat}}

\def\*shl*#1{\c*=\the\n*\d**\advance\c*#1\a**\d*\dt*\advance\d*#1\b**
\a*=\the\toks0\c*\b*=\the\toks1\d*\advance\a*-\b*
\b*=\the\toks1\c*\d*=\the\toks0\d*\advance\b*\d*
\rx* \the\cos*\a* \tmp* \the\sin*\b* \advance\rx*-\tmp*
\ry* \the\cos*\b* \tmp* \the\sin*\a* \advance\ry*\tmp*
\raise\ry*\rlap{\kern\rx*\unhcopy\spl*}}

\def\calcnormal*#1{\b**=10000sp\a**\b**\k*\n*\advance\k*-\m*
\multiply\a**\k*\divide\a**\m*\a**=#1\a**\ifdim\a**<0pt\a**-\a**\fi
\ifdim\a**>\b**\d*=.96\a**\advance\d*.4\b**
\else\d*=.96\b**\advance\d*.4\a**\fi
\d*=.01\d*\r*\d*\divide\a**\r*\divide\b**\r*
\ifnum\k*<0\a**-\a**\fi\d*=#1\d*\ifdim\d*<0pt\b**-\b**\fi
\k*\a**\a**=\the\k*\dd*\k*\b**\b**=\the\k*\dd*}

\def\wavearcto#1(#2,#3)[#4]{\rlap{\toks0={#2}\toks1={#3}\relax
\calcnum*#1(#2,#3)\c*=\the\N*5sp\a*=.4\c*\a*=#4\a*\ifdim\a*<0pt\a*-\a*\fi
\advance\c*\a*\N*\c*\divide\N*20\multiply\N*2\advance\N*-1\multiply\N*4\relax
\d**=#1\Lengthunit\dd*=.012\d**
\divide\dd*\*ths \multiply\dd*\magnitude
\ifdim\d**<0pt\d**-\d**\fi\L*=.25\d**
\divide\d**\N*\divide\dd*\N*\L*=#4\L*\divide\L*\*ths
\m*\N*\divide\m*2\dm*=\the\m*0sp\l*\dm*
\sm*\n*\*one\loop\calcnormal*{#4}\calcparab*
\*shl*{1}\advance\n*\*one\calcparab*
\*shl*{1.3}\advance\n*\*one\calcparab*
\*shl*{1}\advance\n*2\dd*-\dd*\ifnum\n*<\N*\repeat\n*\N*\shl**{0pt}}}

\def\triangarcto#1(#2,#3)[#4]{\rlap{\toks0={#2}\toks1={#3}\relax
\calcnum*#1(#2,#3)\c*=\the\N*5sp\a*=.4\c*\a*=#4\a*\ifdim\a*<0pt\a*-\a*\fi
\advance\c*\a*\N*\c*\divide\N*20\multiply\N*2\advance\N*-1\multiply\N*2\relax
\d**=#1\Lengthunit\dd*=.012\d**
\divide\dd*\*ths \multiply\dd*\magnitude
\ifdim\d**<0pt\d**-\d**\fi\L*=.25\d**
\divide\d**\N*\divide\dd*\N*\L*=#4\L*\divide\L*\*ths
\m*\N*\divide\m*2\dm*=\the\m*0sp\l*\dm*
\sm*\n*\*one\loop\calcnormal*{#4}\calcparab*
\*shl*{1}\advance\n*2\dd*-\dd*\ifnum\n*<\N*\repeat\n*\N*\shl**{0pt}}}

\def\hr*#1{\L*=\xscale\Lengthunit\ifnum
\angle**=0\clap{\vrule width#1\L* height.1pt}\else
\L*=#1\L*\L*=.5\L*\rmov*(-\L*,0pt){\sm*}\rmov*(\L*,0pt){\sl*}\fi}

\def\shade#1[#2]{\rlap{\Lengthunit=#1\Lengthunit
\special{em:linewidth .001pt}\relax
\mov(0,#2.05){\hr*{.994}}\mov(0,#2.1){\hr*{.980}}\relax
\mov(0,#2.15){\hr*{.953}}\mov(0,#2.2){\hr*{.916}}\relax
\mov(0,#2.25){\hr*{.867}}\mov(0,#2.3){\hr*{.798}}\relax
\mov(0,#2.35){\hr*{.715}}\mov(0,#2.4){\hr*{.603}}\relax
\mov(0,#2.45){\hr*{.435}}\special{em:linewidth \the\linwid*}}}

\def\dshade#1[#2]{\rlap{\special{em:linewidth .001pt}\relax
\Lengthunit=#1\Lengthunit\if#2-\def\t*{+}\else\def\t*{-}\fi
\mov(0,\t*.025){\relax
\mov(0,#2.05){\hr*{.995}}\mov(0,#2.1){\hr*{.988}}\relax
\mov(0,#2.15){\hr*{.969}}\mov(0,#2.2){\hr*{.937}}\relax
\mov(0,#2.25){\hr*{.893}}\mov(0,#2.3){\hr*{.836}}\relax
\mov(0,#2.35){\hr*{.760}}\mov(0,#2.4){\hr*{.662}}\relax
\mov(0,#2.45){\hr*{.531}}\mov(0,#2.5){\hr*{.320}}\relax
\special{em:linewidth \the\linwid*}}}}

\def\vdot{\rlap{\kern-1.9pt\lower1.8pt\hbox{$\scriptstyle\bullet$}}}
\def\vtimes{\rlap{\kern-3pt\lower1.8pt\hbox{$\scriptstyle\times$}}}
\def\vDot{\rlap{\kern-2.3pt\lower2.7pt\hbox{$\bullet$}}}
\def\vTimes{\rlap{\kern-3.6pt\lower2.4pt\hbox{$\times$}}}

\def\arc(#1)[#2,#3]{{\k*=#2\l*=#3\m*=\l*
\advance\m*-6\ifnum\k*>\l*\relax\else
{\rotate(#2)\mov(#1,0){\sm*}}\loop
\ifnum\k*<\m*\advance\k*5{\rotate(\k*)\mov(#1,0){\sl*}}\repeat
{\rotate(#3)\mov(#1,0){\sl*}}\fi}}

\def\dasharc(#1)[#2,#3]{{\k**=#2\n*=#3\advance\n*-1\advance\n*-\k**
\L*=1000sp\L*#1\L* \multiply\L*\n* \multiply\L*\Nhalfperiods
\divide\L*57\N*\L* \divide\N*2000\ifnum\N*=0\N*1\fi
\r*\n*  \divide\r*\N* \ifnum\r*<2\r*2\fi
\m**\r* \divide\m**2 \l**\r* \advance\l**-\m** \N*\n* \divide\N*\r*
\k**\r* \multiply\k**\N* \dn*\n*
\advance\dn*-\k** \divide\dn*2\advance\dn*\*one
\r*\l** \divide\r*2\advance\dn*\r* \advance\N*-2\k**#2\relax
\ifnum\l**<6{\rotate(#2)\mov(#1,0){\sm*}}\advance\k**\dn*
{\rotate(\k**)\mov(#1,0){\sl*}}\advance\k**\m**
{\rotate(\k**)\mov(#1,0){\sm*}}\loop
\advance\k**\l**{\rotate(\k**)\mov(#1,0){\sl*}}\advance\k**\m**
{\rotate(\k**)\mov(#1,0){\sm*}}\advance\N*-1\ifnum\N*>0\repeat
{\rotate(#3)\mov(#1,0){\sl*}}\else\advance\k**\dn*
\arc(#1)[#2,\k**]\loop\advance\k**\m** \r*\k**
\advance\k**\l** {\arc(#1)[\r*,\k**]}\relax
\advance\N*-1\ifnum\N*>0\repeat
\advance\k**\m**\arc(#1)[\k**,#3]\fi}}

\def\triangarc#1(#2)[#3,#4]{{\k**=#3\n*=#4\advance\n*-\k**
\L*=1000sp\L*#2\L* \multiply\L*\n* \multiply\L*\Nhalfperiods
\divide\L*57\N*\L* \divide\N*1000\ifnum\N*=0\N*1\fi
\d**=#2\Lengthunit \d*\d** \divide\d*57\multiply\d*\n*
\r*\n*  \divide\r*\N* \ifnum\r*<2\r*2\fi
\m**\r* \divide\m**2 \l**\r* \advance\l**-\m** \N*\n* \divide\N*\r*
\dt*\d* \divide\dt*\N* \dt*.5\dt* \dt*#1\dt*
\divide\dt*1000\multiply\dt*\magnitude
\k**\r* \multiply\k**\N* \dn*\n* \advance\dn*-\k** \divide\dn*2\relax
\r*\l** \divide\r*2\advance\dn*\r* \advance\N*-1\k**#3\relax
{\rotate(#3)\mov(#2,0){\sm*}}\advance\k**\dn*
{\rotate(\k**)\mov(#2,0){\sl*}}\advance\k**-\m**\advance\l**\m**\loop\dt*-\dt*
\d*\d** \advance\d*\dt*
\advance\k**\l**{\rotate(\k**)\rmov*(\d*,0pt){\sl*}}%
\advance\N*-1\ifnum\N*>0\repeat\advance\k**\m**
{\rotate(\k**)\mov(#2,0){\sl*}}{\rotate(#4)\mov(#2,0){\sl*}}}}

\def\wavearc#1(#2)[#3,#4]{{\k**=#3\n*=#4\advance\n*-\k**
\L*=4000sp\L*#2\L* \multiply\L*\n* \multiply\L*\Nhalfperiods
\divide\L*57\N*\L* \divide\N*1000\ifnum\N*=0\N*1\fi
\d**=#2\Lengthunit \d*\d** \divide\d*57\multiply\d*\n*
\r*\n*  \divide\r*\N* \ifnum\r*=0\r*1\fi
\m**\r* \divide\m**2 \l**\r* \advance\l**-\m** \N*\n* \divide\N*\r*
\dt*\d* \divide\dt*\N* \dt*.7\dt* \dt*#1\dt*
\divide\dt*1000\multiply\dt*\magnitude
\k**\r* \multiply\k**\N* \dn*\n* \advance\dn*-\k** \divide\dn*2\relax
\divide\N*4\advance\N*-1\k**#3\relax
{\rotate(#3)\mov(#2,0){\sm*}}\advance\k**\dn*
{\rotate(\k**)\mov(#2,0){\sl*}}\advance\k**-\m**\advance\l**\m**\loop\dt*-\dt*
\d*\d** \advance\d*\dt* \dd*\d** \advance\dd*1.3\dt*
\advance\k**\r*{\rotate(\k**)\rmov*(\d*,0pt){\sl*}}\relax
\advance\k**\r*{\rotate(\k**)\rmov*(\dd*,0pt){\sl*}}\relax
\advance\k**\r*{\rotate(\k**)\rmov*(\d*,0pt){\sl*}}\relax
\advance\k**\r*
\advance\N*-1\ifnum\N*>0\repeat\advance\k**\m**
{\rotate(\k**)\mov(#2,0){\sl*}}{\rotate(#4)\mov(#2,0){\sl*}}}}

\def\gmov*#1(#2,#3)#4{\rlap{\L*=#1\Lengthunit
\xL*=#2\L* \yL*=#3\L*
\rx* \gcos*\xL* \tmp* \gsin*\yL* \advance\rx*-\tmp*
\ry* \gcos*\yL* \tmp* \gsin*\xL* \advance\ry*\tmp*
\rx*=\xscale\rx* \ry*=\yscale\ry*
\xL* \the\cos*\rx* \tmp* \the\sin*\ry* \advance\xL*-\tmp*
\yL* \the\cos*\ry* \tmp* \the\sin*\rx* \advance\yL*\tmp*
\kern\xL*\raise\yL*\hbox{#4}}}

\def\rgmov*(#1,#2)#3{\rlap{\xL*#1\yL*#2\relax
\rx* \gcos*\xL* \tmp* \gsin*\yL* \advance\rx*-\tmp*
\ry* \gcos*\yL* \tmp* \gsin*\xL* \advance\ry*\tmp*
\rx*=\xscale\rx* \ry*=\yscale\ry*
\xL* \the\cos*\rx* \tmp* \the\sin*\ry* \advance\xL*-\tmp*
\yL* \the\cos*\ry* \tmp* \the\sin*\rx* \advance\yL*\tmp*
\kern\xL*\raise\yL*\hbox{#3}}}

\def\Earc(#1)[#2,#3][#4,#5]{{\k*=#2\l*=#3\m*=\l*
\advance\m*-6\ifnum\k*>\l*\relax\else\def\xscale{#4}\def\yscale{#5}\relax
{\angle**0\rotate(#2)}\gmov*(#1,0){\sm*}\loop
\ifnum\k*<\m*\advance\k*5\relax
{\angle**0\rotate(\k*)}\gmov*(#1,0){\sl*}\repeat
{\angle**0\rotate(#3)}\gmov*(#1,0){\sl*}\relax
\def\xscale{1}\def\yscale{1}\fi}}

\def\dashEarc(#1)[#2,#3][#4,#5]{{\k**=#2\n*=#3\advance\n*-1\advance\n*-\k**
\L*=1000sp\L*#1\L* \multiply\L*\n* \multiply\L*\Nhalfperiods
\divide\L*57\N*\L* \divide\N*2000\ifnum\N*=0\N*1\fi
\r*\n*  \divide\r*\N* \ifnum\r*<2\r*2\fi
\m**\r* \divide\m**2 \l**\r* \advance\l**-\m** \N*\n* \divide\N*\r*
\k**\r*\multiply\k**\N* \dn*\n* \advance\dn*-\k** \divide\dn*2\advance\dn*\*one
\r*\l** \divide\r*2\advance\dn*\r* \advance\N*-2\k**#2\relax
\ifnum\l**<6\def\xscale{#4}\def\yscale{#5}\relax
{\angle**0\rotate(#2)}\gmov*(#1,0){\sm*}\advance\k**\dn*
{\angle**0\rotate(\k**)}\gmov*(#1,0){\sl*}\advance\k**\m**
{\angle**0\rotate(\k**)}\gmov*(#1,0){\sm*}\loop
\advance\k**\l**{\angle**0\rotate(\k**)}\gmov*(#1,0){\sl*}\advance\k**\m**
{\angle**0\rotate(\k**)}\gmov*(#1,0){\sm*}\advance\N*-1\ifnum\N*>0\repeat
{\angle**0\rotate(#3)}\gmov*(#1,0){\sl*}\def\xscale{1}\def\yscale{1}\else
\advance\k**\dn* \Earc(#1)[#2,\k**][#4,#5]\loop\advance\k**\m** \r*\k**
\advance\k**\l** {\Earc(#1)[\r*,\k**][#4,#5]}\relax
\advance\N*-1\ifnum\N*>0\repeat
\advance\k**\m**\Earc(#1)[\k**,#3][#4,#5]\fi}}

\def\triangEarc#1(#2)[#3,#4][#5,#6]{{\k**=#3\n*=#4\advance\n*-\k**
\L*=1000sp\L*#2\L* \multiply\L*\n* \multiply\L*\Nhalfperiods
\divide\L*57\N*\L* \divide\N*1000\ifnum\N*=0\N*1\fi
\d**=#2\Lengthunit \d*\d** \divide\d*57\multiply\d*\n*
\r*\n*  \divide\r*\N* \ifnum\r*<2\r*2\fi
\m**\r* \divide\m**2 \l**\r* \advance\l**-\m** \N*\n* \divide\N*\r*
\dt*\d* \divide\dt*\N* \dt*.5\dt* \dt*#1\dt*
\divide\dt*1000\multiply\dt*\magnitude
\k**\r* \multiply\k**\N* \dn*\n* \advance\dn*-\k** \divide\dn*2\relax
\r*\l** \divide\r*2\advance\dn*\r* \advance\N*-1\k**#3\relax
\def\xscale{#5}\def\yscale{#6}\relax
{\angle**0\rotate(#3)}\gmov*(#2,0){\sm*}\advance\k**\dn*
{\angle**0\rotate(\k**)}\gmov*(#2,0){\sl*}\advance\k**-\m**
\advance\l**\m**\loop\dt*-\dt* \d*\d** \advance\d*\dt*
\advance\k**\l**{\angle**0\rotate(\k**)}\rgmov*(\d*,0pt){\sl*}\relax
\advance\N*-1\ifnum\N*>0\repeat\advance\k**\m**
{\angle**0\rotate(\k**)}\gmov*(#2,0){\sl*}\relax
{\angle**0\rotate(#4)}\gmov*(#2,0){\sl*}\def\xscale{1}\def\yscale{1}}}

\def\waveEarc#1(#2)[#3,#4][#5,#6]{{\k**=#3\n*=#4\advance\n*-\k**
\L*=4000sp\L*#2\L* \multiply\L*\n* \multiply\L*\Nhalfperiods
\divide\L*57\N*\L* \divide\N*1000\ifnum\N*=0\N*1\fi
\d**=#2\Lengthunit \d*\d** \divide\d*57\multiply\d*\n*
\r*\n*  \divide\r*\N* \ifnum\r*=0\r*1\fi
\m**\r* \divide\m**2 \l**\r* \advance\l**-\m** \N*\n* \divide\N*\r*
\dt*\d* \divide\dt*\N* \dt*.7\dt* \dt*#1\dt*
\divide\dt*1000\multiply\dt*\magnitude
\k**\r* \multiply\k**\N* \dn*\n* \advance\dn*-\k** \divide\dn*2\relax
\divide\N*4\advance\N*-1\k**#3\def\xscale{#5}\def\yscale{#6}\relax
{\angle**0\rotate(#3)}\gmov*(#2,0){\sm*}\advance\k**\dn*
{\angle**0\rotate(\k**)}\gmov*(#2,0){\sl*}\advance\k**-\m**
\advance\l**\m**\loop\dt*-\dt*
\d*\d** \advance\d*\dt* \dd*\d** \advance\dd*1.3\dt*
\advance\k**\r*{\angle**0\rotate(\k**)}\rgmov*(\d*,0pt){\sl*}\relax
\advance\k**\r*{\angle**0\rotate(\k**)}\rgmov*(\dd*,0pt){\sl*}\relax
\advance\k**\r*{\angle**0\rotate(\k**)}\rgmov*(\d*,0pt){\sl*}\relax
\advance\k**\r*
\advance\N*-1\ifnum\N*>0\repeat\advance\k**\m**
{\angle**0\rotate(\k**)}\gmov*(#2,0){\sl*}\relax
{\angle**0\rotate(#4)}\gmov*(#2,0){\sl*}\def\xscale{1}\def\yscale{1}}}

\newcount\CatcodeOfAtSign
\CatcodeOfAtSign=\the\catcode`\@
\catcode`\@=11
\def\@arc#1[#2][#3]{\rlap{\Lengthunit=#1\Lengthunit
\sm*\l*arc(#2.1914,#3.0381)[#2][#3]\relax
\mov(#2.1914,#3.0381){\l*arc(#2.1622,#3.1084)[#2][#3]}\relax
\mov(#2.3536,#3.1465){\l*arc(#2.1084,#3.1622)[#2][#3]}\relax
\mov(#2.4619,#3.3086){\l*arc(#2.0381,#3.1914)[#2][#3]}}}

\def\dash@arc#1[#2][#3]{\rlap{\Lengthunit=#1\Lengthunit
\d*arc(#2.1914,#3.0381)[#2][#3]\relax
\mov(#2.1914,#3.0381){\d*arc(#2.1622,#3.1084)[#2][#3]}\relax
\mov(#2.3536,#3.1465){\d*arc(#2.1084,#3.1622)[#2][#3]}\relax
\mov(#2.4619,#3.3086){\d*arc(#2.0381,#3.1914)[#2][#3]}}}

\def\wave@arc#1[#2][#3]{\rlap{\Lengthunit=#1\Lengthunit
\w*lin(#2.1914,#3.0381)\relax
\mov(#2.1914,#3.0381){\w*lin(#2.1622,#3.1084)}\relax
\mov(#2.3536,#3.1465){\w*lin(#2.1084,#3.1622)}\relax
\mov(#2.4619,#3.3086){\w*lin(#2.0381,#3.1914)}}}

\def\bezier#1(#2,#3)(#4,#5)(#6,#7){\N*#1\l*\N* \advance\l*\*one
\d* #4\Lengthunit \advance\d* -#2\Lengthunit \multiply\d* \*two
\b* #6\Lengthunit \advance\b* -#2\Lengthunit
\advance\b*-\d* \divide\b*\N*
\d** #5\Lengthunit \advance\d** -#3\Lengthunit \multiply\d** \*two
\b** #7\Lengthunit \advance\b** -#3\Lengthunit
\advance\b** -\d** \divide\b**\N*
\mov(#2,#3){\sm*{\loop\ifnum\m*<\l*
\a*\m*\b* \advance\a*\d* \divide\a*\N* \multiply\a*\m*
\a**\m*\b** \advance\a**\d** \divide\a**\N* \multiply\a**\m*
\rmov*(\a*,\a**){\unhcopy\spl*}\advance\m*\*one\repeat}}}

\catcode`\*=12

\newcount\n@ast

\def\n@ast@#1{\n@ast0\relax\get@ast@#1\end}
\def\get@ast@#1{\ifx#1\end\let\next\relax\else
\ifx#1*\advance\n@ast1\fi\let\next\get@ast@\fi\next}

\newif\if@up \newif\if@dwn
\def\up@down@#1{\@upfalse\@dwnfalse
\if#1u\@uptrue\fi\if#1U\@uptrue\fi\if#1+\@uptrue\fi
\if#1d\@dwntrue\fi\if#1D\@dwntrue\fi\if#1-\@dwntrue\fi}

\def\halfcirc#1(#2)[#3]{{\Lengthunit=#2\Lengthunit\up@down@{#3}\relax
\if@up\mov(0,.5){\@arc[-][-]\@arc[+][-]}\fi
\if@dwn\mov(0,-.5){\@arc[-][+]\@arc[+][+]}\fi
\def\lft{\mov(0,.5){\@arc[-][-]}\mov(0,-.5){\@arc[-][+]}}\relax
\def\rght{\mov(0,.5){\@arc[+][-]}\mov(0,-.5){\@arc[+][+]}}\relax
\if#3l\lft\fi\if#3L\lft\fi\if#3r\rght\fi\if#3R\rght\fi
\n@ast@{#1}\relax
\ifnum\n@ast>0\if@up\shade[+]\fi\if@dwn\shade[-]\fi\fi
\ifnum\n@ast>1\if@up\dshade[+]\fi\if@dwn\dshade[-]\fi\fi}}

\def\halfdashcirc(#1)[#2]{{\Lengthunit=#1\Lengthunit\up@down@{#2}\relax
\if@up\mov(0,.5){\dash@arc[-][-]\dash@arc[+][-]}\fi
\if@dwn\mov(0,-.5){\dash@arc[-][+]\dash@arc[+][+]}\fi
\def\lft{\mov(0,.5){\dash@arc[-][-]}\mov(0,-.5){\dash@arc[-][+]}}\relax
\def\rght{\mov(0,.5){\dash@arc[+][-]}\mov(0,-.5){\dash@arc[+][+]}}\relax
\if#2l\lft\fi\if#2L\lft\fi\if#2r\rght\fi\if#2R\rght\fi}}

\def\halfwavecirc(#1)[#2]{{\Lengthunit=#1\Lengthunit\up@down@{#2}\relax
\if@up\mov(0,.5){\wave@arc[-][-]\wave@arc[+][-]}\fi
\if@dwn\mov(0,-.5){\wave@arc[-][+]\wave@arc[+][+]}\fi
\def\lft{\mov(0,.5){\wave@arc[-][-]}\mov(0,-.5){\wave@arc[-][+]}}\relax
\def\rght{\mov(0,.5){\wave@arc[+][-]}\mov(0,-.5){\wave@arc[+][+]}}\relax
\if#2l\lft\fi\if#2L\lft\fi\if#2r\rght\fi\if#2R\rght\fi}}

\catcode`\*=11

\def\Circle#1(#2){\halfcirc#1(#2)[u]\halfcirc#1(#2)[d]\n@ast@{#1}\relax
\ifnum\n@ast>0\L*=\xscale\Lengthunit
\ifnum\angle**=0\clap{\vrule width#2\L* height.1pt}\else
\L*=#2\L*\L*=.5\L*\special{em:linewidth .001pt}\relax
\rmov*(-\L*,0pt){\sm*}\rmov*(\L*,0pt){\sl*}\relax
\special{em:linewidth \the\linwid*}\fi\fi}

\catcode`\*=12

\def\wavecirc(#1){\halfwavecirc(#1)[u]\halfwavecirc(#1)[d]}
\def\dashcirc(#1){\halfdashcirc(#1)[u]\halfdashcirc(#1)[d]}

\def\xscale{1}

\def\yscale{1}

\def\Ellipse#1(#2)[#3,#4]{\def\xscale{#3}\def\yscale{#4}\relax
\Circle#1(#2)\def\xscale{1}\def\yscale{1}}

\def\dashEllipse(#1)[#2,#3]{\def\xscale{#2}\def\yscale{#3}\relax
\dashcirc(#1)\def\xscale{1}\def\yscale{1}}

\def\waveEllipse(#1)[#2,#3]{\def\xscale{#2}\def\yscale{#3}\relax
\wavecirc(#1)\def\xscale{1}\def\yscale{1}}

\def\halfEllipse#1(#2)[#3][#4,#5]{\def\xscale{#4}\def\yscale{#5}\relax
\halfcirc#1(#2)[#3]\def\xscale{1}\def\yscale{1}}

\def\halfdashEllipse(#1)[#2][#3,#4]{\def\xscale{#3}\def\yscale{#4}\relax
\halfdashcirc(#1)[#2]\def\xscale{1}\def\yscale{1}}

\def\halfwaveEllipse(#1)[#2][#3,#4]{\def\xscale{#3}\def\yscale{#4}\relax
\halfwavecirc(#1)[#2]\def\xscale{1}\def\yscale{1}}

\catcode`\@=\the\CatcodeOfAtSign

\title{On the higher-derivative supersymmetric gauge theory}

\author{F. S. Gama}
\email{fgama@fisica.ufpb.br}
\affiliation{Departamento de F\'{\i}sica, Universidade Federal da 
Para\'{\i}ba\\
 Caixa Postal 5008, 58051-970, Jo\~ao Pessoa, Para\'{\i}ba, Brazil}

\author{M. Gomes}
\email{mgomes@fma.if.usp.br}
\affiliation{Departamento de F\'{\i}sica Matematica, Instituto de
  F\'{\i}sica,
\\
Universidade de S\~ao Paulo,
 Caixa Postal 66318, S\~ao Paulo, Brazil}

\author{J. R. Nascimento}
\email{jroberto@fisica.ufpb.br}
\affiliation{Departamento de F\'{\i}sica, Universidade Federal da 
Para\'{\i}ba\\
 Caixa Postal 5008, 58051-970, Jo\~ao Pessoa, Para\'{\i}ba, Brazil}

\author{A. Yu. Petrov}
\email{petrov@fisica.ufpb.br}
\affiliation{Departamento de F\'{\i}sica, Universidade Federal da 
Para\'{\i}ba\\
 Caixa Postal 5008, 58051-970, Jo\~ao Pessoa, Para\'{\i}ba, Brazil}

\author{A. J. da Silva}
\email{ajsilva@fma.if.usp.br}
\affiliation{Departamento de F\'{\i}sica Matematica, Instituto de
  F\'{\i}sica,
\\
Universidade de S\~ao Paulo,
 Caixa Postal 66318, S\~ao Paulo, Brazil}

\begin{abstract}
We study the one-loop low-energy effective action for the
higher-derivative superfield gauge theory coupled to a chiral matter.
\end{abstract}
\maketitle

\section{Introduction}

The use of higher-derivatives has been proposed as a way to tame the ultraviolet behavior of physically relevant models. Actually, a finite version of QED was put forward by Lee and Wick about fourty years ago \cite{Lee}; that proposal was nevertheless beset by the presence of spurious degrees of freedom which  induce indefinite metric in the space of states jeopardizing unitarity and so required special treatment. After a long story, recently the idea was revived in the so-called Lee-Wick standard model leading to some new insights in the hierachy problem \cite{Wise}. Furthermore, following a similar thread, higher-curvature gravitational models were considered; in spite of  the possible breaking of unitarity, they furnish  renormalizable quantum gravity models \cite{Stelle} which may be useful for cosmological applications (for recent review work in this direction see
\cite{Faraoni}).

 In
supersymmetric models, a higher-derivative regularization method was
proposed in \cite{Ili}. Further, interest in the higher-derivative
supersymmetric field theories increased not only due to their
application within the regularization context (see \cite{Step} for
many examples) but also in other contexts. For example, the
higher-derivative supergravity model, which, in the superconformal
sector, can be treated as a natural higher-derivative generalization
of the Wess-Zumino model, has been studied in \cite{bp1,bp2}. Afterwards,
some classical aspects of the very generic class of the
higher-derivative chiral superfield models have been considered in
\cite{Ant}, and the lower perturbative corrections in such theories
were obtained in 
\cite{ourhigh}.

Therefore, the natural continuation of these studies could be the
construction of a consistent higher-derivative gauge theory coupled to
a chiral matter. Within this paper, we construct such a model
(unfortunately, this construction is probably
possible only in the Abelian case). We calculate the one-loop
low-energy effective action for such a theory generalizing the results
of \cite{YMEP} where the usual super-Yang-Mills field coupled to a chiral
matter has been studied in the one-loop approximation. Within this
study, we employ the superfield approach for calculating the
supersymmetric effective potential developed in Refs. \cite{efpot,BCP,bp3}.

\section{Effective action in higher-derivative superfield gauge  
theories:
  general  approach}

Let us formulate the supersymmetric higher-derivative gauge theory. It
is well known \cite{BK0} that under the usual gauge transformation of
the gauge  superfield $v$
\bea
\label{tra0}
e^{gv}\to e^{-ig\bar{\Lambda}}e^{gv}e^{ig\Lambda},
\eea
where $\Lambda$ is a chiral parameter, and $\bar{\Lambda}$ is an
antichiral  one, the superfield strength
\bea
W_{\alpha}=\bar{D}^2(e^{-gv}D_{\alpha}e^{gv}),
\eea
 is transformed as
\bea
\label{gtra}
W_{\alpha}\to e^{-ig\Lambda}W_{\alpha}e^{ig\Lambda}.
\eea
In the Abelian case, the strength $W_{\alpha}$ is invariant.

Now, let us try to implement the higher derivatives in the
super-Yang-Mills  theory with matter whose standard form is
\bea
S=\int d^8z \bar{\phi}e^{gV}\phi+\frac{1}{2g^2}{\rm tr}\int
d^6zW^{\alpha} W_{\alpha}.
\eea
In the non-Abelian case, the $\phi$ is a column vector (we can
consider as well the  case when the $\phi$ is not an isospinor but
also the  Lie-algebra valued field, as $W^{\alpha}$ is; however, it is
not relevant within our approach).
The first idea consists in inserting of some operator in the purely
gauge part  of the action rewriting it as
\bea
S_W=\frac{1}{2g^2}{\rm tr}\int d^6zW^{\alpha}{\cal O}W_{\alpha}.
\eea
In the non-Abelian case, this action is invariant under these gauge
transformations if and only if the operator ${\cal O}$ satisfies the condition
\bea
e^{ig\Lambda}{\cal O}e^{-ig\Lambda}={\cal O}.
\eea
Expanding this equation order by order in $g$, we find that this
condition can be satisfied only if $[\Lambda,{\cal O}]=0$, i.e. the
operator ${\cal O}$ is proportional to the unit operator, and neither
spinor, nor vector derivatives are allowed. Therefore, in the
non-Abelian case, higher derivatives cannot be introduced in a manner
compatible with the gauge  invariance.

So, we will restrict ourselves to develop the Abelian case where the
gauge transformations  are reduced to \cite{BK0}
\bea
v\to v+i(\Lambda-\bar{\Lambda}),
\eea
and the strength $W_{\alpha}$ is explicitly gauge invariant. In this
case, as a first, simplest way, we can write the following
higher-derivative action for the Abelian gauge  superfield:
\bea
S_W=\frac{1}{2g^2}\int
d^6zW^{\alpha}(\Box+m^2)W_{\alpha}=-\frac{1}{16}\int  d^8 z v
D^{\alpha} \bar{D}^2D_{\alpha}(\Box+m^2)v.
\eea
We note that this action (but with $m=0$) was earlier \cite{bp2}
employed as an auxiliary tool for studying the higher-derivative
generalization of the Wess-Zumino model. However, the detailed studies
of the properties of this action were not carried out. Now, we should
develop now its consistent coupling to the chiral  matter.
It is clear that this coupling is invariant if we suppose that the
chiral and antichiral  superfields are transformed as
\bea
\phi\to e^{-i\Lambda}\phi,\quad\, \bar{\phi}\to \bar{\phi}e^{i\bar{\Lambda}}.
\eea
A natural generalization of the chiral superfield action
\bea
S_{\Phi}=\int d^8z \bar{\phi}e^{gV}{\cal R}\phi,
\eea
with ${\cal R}$ some operator, is gauge invariant again, if ${\cal R}$
is proportional to the unit operator, both in the Abelian and in the
non-Abelian case. Therefore, the simplest action for the
higher-derivative supersymmetric gauge theory, that is, for the
supersymmetric scalar QED is
\bea
\label{action}
S=\int d^8z \bar{\phi}e^{gV}\phi-\frac{1}{16}\int d^8 z v
D^{\alpha}\bar{D}^2 D_{\alpha}(\Box+m^2)v.
\eea
In principle, this action can be generalized by introducing the some
self-couplings of set of the chiral superfields (see
f.e. \cite{Kuz}), 
however, we here want to study the generic structure for the one-loop
low-energy effective action for the chiral  superfield.

To fix the gauge, we add the gauge-fixing term:
\bea
S_{gf}=\frac{1}{16\alpha}\int d^8z v D^2\bar{D}^2(\Box+m^2)v.
\eea
Here $\alpha$ is the gauge-fixing parameter.

Following \cite{BK0}, the low-energy effective action in the theory of
chiral scalar superfield is described by the K\"{a}hlerian effective
potential which depends only on chiral and antichiral superfields but
not on their derivatives. Namely this effective potential will be the
principal object of study in the paper. We note that since the model
(\ref{action}) does not involve chiral self-coupling of the matter
fields, the chiral effective potential in it will be identically equal
to zero unlike in the Wess-Zumino model and other models where such a
coupling is present. The ghosts are completely factorized since the
theory is Abelian.

The propagators in the theory (\ref{action}) are very similar to the 
propagators in the usual super-Yang-Mills theory \cite{BK0}:
\bea
&&<\phi(z_1)\bar{\phi}(z_2)>=\frac{\bar{D}^2D^2}{16\Box}\delta^8(z_1-z_2);\,
<\bar{\phi}(z_1)\phi(z_2)>=\frac{D^2\bar{D}^2}{16\Box}\delta^8(z_1-z_2);
\nonumber\\
&&<v(z_1)v(z_2)>=-\frac{1}{\Box(\Box+m^2)}
(-\frac{D^{\alpha}\bar{D}^2D_{\alpha}}{8\Box}+\alpha\frac{\{\bar{D}^2,D^2\}}{16
\Box})
\delta^8(z_1-z_2).
\eea
It is convenient to express these propagators in terms of the
projecting 
operators \cite{BK0}
$$
\Pi_0=\frac{\{\bar{D}^2,D^2\}}{16\Box},\quad\,\Pi_{1/2}=-\frac{D^{\alpha}
  \bar{D}^2D_{\alpha}}{8\Box}.
$$
Indeed, it is clear that $\Pi_0^n=\Pi_0$, $\Pi_{1/2}^n=\Pi_{1/2}$ (for
any  integer $n\geq 1$), $\Pi_0\Pi_{1/2}=\Pi_{1/2}\Pi_0=0$. Thus, we can write
\bea
&&<\phi(z_1)\bar{\phi}(z_2)>+<\bar{\Phi}(z_1)\Phi(z_2)>=\Pi_0\delta^8(z_1-z_2);
\nonumber\\
&&<v(z_1)v(z_2)>=-\frac{1}{\Box(\Box+m^2)}(\Pi_{1/2}+\alpha\Pi_0)\delta^8(z_1-
z_2).
\eea
Here we emphasized the combination
$<\Phi(z_1)\bar{\Phi}(z_2)>+<\bar{\Phi}(z_1)\Phi(z_2)>$ since namely
it will arise in many cases including the contributions to the
one-loop 
k\"{a}hlerian effective potential.

\section{One-loop calculations}

Now, let us start with study of the one-loop k\"{a}hlerian
potential. At the one-loop order, we will have two types of
contributions. In the first of them, all diagrams involve only the
gauge field  propagators:

\vspace*{2mm}

\hspace{2.0cm}
\Lengthunit=1.2cm
\Linewidth{.5pt}
\GRAPH(hsize=4){\mov(.5,0){\wavecirc(1)}\mov(.5,.5){\lin(-.5,.5)\lin(.5,.5)}
\mov(3,0){\mov(.5,0){\wavecirc(1)}\mov(.5,.5){\lin(-.5,.5)\lin(.5,.5)}
\mov(.5,-.5){\lin(-.5,-.5)\lin(.5,-.5)}}
\mov(6,0){\mov(.5,0){\wavecirc(1)}\mov(.5,.5){\lin(-.5,.5)\lin(.5,.5)}
\mov(.5,-.5){\lin(-.5,-.5)\lin(.5,-.5)}\mov(.9,0){\lin(.5,-.5)\lin(.5,.5)}}
\ind(80,0){\ldots}
}

\vspace*{2mm}

The contribution of the sum of these diagrams can be expressed as
\bea
K^{(1)}_a=\int
d^8z_1\sum\limits_{n=1}^{\infty}\frac{(-1)^n}{2n}(g^2\Phi\bar{\Phi}
\frac{1}{\Box(\Box+m^2)}(\Pi_{1/2}+\alpha\Pi_0))^n\delta_{12}|_
{\theta_1=\theta_2},
\eea
where $\frac{1}{n}$ is a symmetry factor. The $\Phi$, $\bar{\Phi}$ are
the background fields.
These diagrams do not involve the triple vertices, only the quartic ones.

Using the properties of the projecting operators, we can write
\bea
K^{(1)}_a=\int
d^8z_1\sum\limits_{n=1}^{\infty}\frac{(-1)^n}{2n}(g^2\Phi\bar{\Phi}
\frac{1}{\Box(\Box+m^2)})^n(\Pi_{1/2}+\alpha^n\Pi_0))\delta_{12}|_{\theta_1=
\theta_2}
.
\eea
Since $\frac{D^2\bar{D}^2}{16}\delta_{12}=1$, we have
$\Box\Pi_0\delta_{12}|_{\theta_1=\theta_2}=2$,  and $\Box\Pi_{1/2} 
\delta_{12}|_{\theta_1=\theta_2}=-2$.  Thus, we have
\bea
K^{(1)}_a=\int
d^8z_1\sum\limits_{n=1}^{\infty}\frac{(-1)^n}{n}\frac{1}{\Box}(g^2\Phi
\bar{\Phi}\frac{1}{\Box(\Box+m^2)})^n(1-\alpha^n)\delta^4(x_1-x_2)|_{x_1=x_2
}.
\eea
By carrying out the Fourier transform $\Box\to -k^2$ and Wick rotation
$k_0=ik_{0E}$  which yields $k^2=k^2_E$, we arrive at
\bea
K^{(1)}_a=-i\int
d^8z\int\frac{d^4k_E}{(2\pi)^4}\sum\limits_{n=1}^{\infty}\frac{(-1)^n}{n}
\frac{1}{\Box}(\frac{g^2\Phi\bar{\Phi}}{k^2(k^2+m^2)})^n
(1-\alpha^n).
\eea
Then, by using the expansion
\bea
\sum\limits_{n=1}^{\infty}\frac{(-a)^n}{n}=-\ln(1+a),
\eea
we have
\bea
\label{k1a}
K^{(1)}_a=i\int
d^8z\int\frac{d^4k_E}{(2\pi)^4}\frac{1}{k^2}\Big[\ln(1+\frac{g^2\Phi\bar{\Phi}
}{k^2(k^2+m^2)})-\ln(1+\frac{\alpha
  g^2\Phi\bar{\Phi}}{k^2(k^2+m^2)}) \Big].
\eea
Notice that at $\alpha=0$ (Landau gauge), the second term in this
expression  vanishes.

The second type of diagrams involves the triple vertices as well. 
We should first introduce a "dressed" propagator

\hspace{4.0cm}
\Lengthunit=1.2cm
\GRAPH(hsize=4){\Linewidth{1.5pt}\wavelin(1,0)\ind(12,0){=}\Linewidth{0.5pt}
  \mov(1.5,0){\wavelin(1,0)}\ind(28,0){+}\mov(3,0){\wavelin(2,0)
\mov(1,0){\lin(-.5,.5)\lin(.5,.5)}}\ind(53,0){+}\ind(58,0){\ldots}
}

In this propagator, the summation over all quartic vertices is
performed. As a  result, this "dressed" propagator is equal to
\bea
<vv>_D&=&<vv>(1+g^2\Phi\bar{\Phi}<vv>+(g^2\Phi\bar{\Phi}<vv>)^2+\ldots)=
\nonumber\\&=&-
\sum\limits_{n=0}^{\infty}(g^2\Phi\bar{\Phi})^n
\frac{1}{(\Box(\Box+m^2))^{n+1}}(\Pi_{1/2}+\alpha\Pi_0)^{n+1}.
\eea
Summing up, we arrive at
\bea
<vv>_D=-(\frac{1}{\Box(\Box+m^2)+g^2\Phi\bar{\Phi}}\Pi_{1/2}+
\frac{\alpha}{\Box(\Box+m^2)+\alpha g^2\Phi\bar{\Phi}} \Pi_0)\delta^8(z_1-z_2).
\eea

As a result, we should sum over diagrams representing themselves as
cycles of  all possible number of links each of which has the form

\vspace*{2mm}

\hspace{6.0cm}
\Lengthunit=2cm
\GRAPH(hsize=2){\Linewidth{1.6pt}\mov(.1,0){\wavelin(1,0)}\Linewidth{.4pt}\mov(.1,0){\lin(-.5,0)}
\lin(0,.7)\mov(1.03,0){\lin(0,.7)}\ind(12,0){\ldots}
}

\vspace*{2mm}

Such diagrams look like

\vspace*{2mm}

\hspace{6.0cm}
\Lengthunit=1cm
\GRAPH(hsize=2){\Linewidth{1.6pt}\halfwavecirc(2)[u]\Linewidth{.4pt}
\mov(-.1,0){\halfcirc(2)[d]\mov(-1,0){\lin(-.5,0)}\mov(1,0){\lin(.5,0)}}
}
\GRAPH(hsize=2){\mov(0,1){
\mov(.12,0){\Linewidth{1.6pt}\wavelin(2,0)\mov(0,-2){\wavelin(2,0)}}
\Linewidth{.4pt}\lin(-.5,.5)\lin(0,-2)\mov(0,-2){\lin(-.5,-.5)}
\mov(1.9,0){\lin(.5,.5)\lin(0,-2)}\mov(1.9,-2){\lin(.5,-.5)}\ind(30,-10){\ldots}
}
}

\vspace*{2mm}

The complete contribution of all these cycles looks like
\bea
K^{(1)}_b=\int
d^8z_1\sum\limits_{n=1}^{\infty}\frac{1}{2n}(g^2\Phi\bar{\Phi}
(<\phi\bar{\phi}>+<\bar{\phi}\phi>)<vv>_D)^n
\delta_{12}|_{\theta_1=\theta_2},
\eea
or, as is the same,
\bea
K^{(1)}_b=\int
d^8z_1\sum\limits_{n=1}^{\infty}\frac{1}{2n}(g^2\Phi\bar{\Phi}
\Pi_0<vv>_D)^n
\delta_{12}|_{\theta_1=\theta_2}.
\eea
By noting that
\bea
\Pi_0 <vv>_D=-\frac{\alpha}{\Box(\Box+m^2)+\alpha g^2\Phi\bar{\Phi}}\Pi_0,
\eea
we can rewrite the expression above as
\bea
K^{(1)}_b=\int d^8z_1\sum\limits_{n=1}^{\infty}\frac{(-1)^n}{2n}(
\frac{\alpha g^2\Phi\bar{\Phi}}{\Box(\Box+m^2)+g^2\Phi\bar{\Phi}}
)^n \Pi_0
\delta_{12}|_{\theta_1=\theta_2}.
\eea
Since $\Box\Pi_0\delta_{12}|_{\theta_1=\theta_2}=2$, we have
\bea
K^{(1)}_b=\int
d^8z_1\sum\limits_{n=1}^{\infty}\frac{1}{n}\frac{1}{\Box}(-\frac{\alpha
  g^2\Phi\bar{\Phi}}{\Box(\Box+m^2)+g^2\Phi\bar{\Phi}})^n\delta^4 
(x_1-x_2)|_{x_1=x_2}.
\eea
Carrying out the Fourier transform, Wick rotation and summation as
above, we 
arrive at
\bea
K^{(1)}_b=i\int d^8z\int\frac{d^4k_E}{(2\pi)^4}\frac{1}{k^2_E}\Big[
\ln(1+\frac{\alpha g^2\Phi\bar{\Phi}}{k^2_E(k^2_E+m^2)})\Big].
\eea
Summing this contribution with $K^{(1)}_a$ (\ref{k1a}), we see that
the $\alpha$ dependent contribution vanishes, and the total one-loop
k\"{a}hlerian effective potential is gauge independent, being, after
returning to the Minkowski space, equal to
\bea
\label{k1a1}
K^{(1)}=\int d^8z\int\frac{d^4k_E}{(2\pi)^4}\frac{1}{k^2_E}\ln\Big[1+
\frac{g^2\Phi\bar{\Phi}}{k^2_E(k^2_E+m^2)}\Big].
\eea
This integral, after change $k^2=u$ and $d^4k=\pi^2udu$, can be
rewritten in 
spherical coordinates
\bea
K^{(1)}=\int d^8z\int_{0}^{\infty}\frac{du}{(4\pi)^2}\ln\Big[1+
\frac{g^2\Phi\bar{\Phi}}{u(u+m^2)}\Big].
\eea
Let us calculate this integral. Although it can be found
straightforwardly, it explicit form seems to be rather
cumbersome. Thertefore it is instructive to estimate its value in some
approximate 
situations.
First, we suppose that the mass is small but non-zero. In this case,
this 
integral can be approximately represented as
\bea
K^{(1)}=\int d^8z\int_{m^2}^{\infty}\frac{du}{(4\pi)^2}\ln\Big[1+
\frac{g^2\Phi\bar\Phi}{u^2}\Big],
\eea
which yields
 \bea
 \label{potential 1}
K^{(1)}=\frac{1}{(4\pi)^2}\int d^8z
\Big\{(g^2\Phi\bar\Phi)^{1/2}\Big[\pi-
2\arctan\Big(\frac{m^2}{(g^2\Phi\bar\Phi)^{1/2}}\Big)\Big]- 
\frac{m^2}{2}\ln\Big(1+\frac{g^2\Phi\bar\Phi}{m^4}\Big)\Big\}.
\eea
Second, we suppose that the mass is zero. In this case we find
 \bea
 \label{potential 2}
K^{(1)}=\frac{1}{16\pi}\int d^8z(g^2\Phi\bar\Phi)^{1/2}.
\eea
It is worth to notice that both results (\ref{potential 1}) and
(\ref{potential 2}) are finite and do not need any renormalization. In
other words, the one-loop k\"{a}hlerian effective potential to the
higher-derivative supersymmetric abelian gauge theory does not display
any divergences, unlike the usual gauge theories 
\cite{YMEP}.

\section{Summary}

We have explicitly found the one-loop k\"{a}hlerian effective
potential in the supersymmetric higher-derivative QED. This term
evidently dominates in the low-energy limit. We note that the chiral
contributions to the effective action, which are known to be typical
for the Wess-Zumino model and its straightforward many-field
generalizations \cite{efpot}, simply do not arise due to the absence of
the chiral self-interaction in the classical action. However, had we
introduced such a term, it is most probably that the chiral
contributions to the effective action could either display the
infrared singularities due to the augmented degree of momenta in the
denominator  (compare with \cite{West} for the usual supersymmetric
QED), in the case when the higher-derivative theory is massless, or
give zero result in the opposite case (one should remind that, in the
Wess-Zumino model, the chiral effective potential also does not emerge
in the massive case). The natural continuation of this study could
consist in the formulation and study of  more generic higher-derivative
supersymmetric gauge models. We will return to this problem in a forthcoming 
paper.

{\bf Acknowledgements.} This work was partially supported by Conselho
Nacional de Desenvolvimento Cient\'{\i}fico e Tecnol\'{o}gico (CNPq)
and Funda\c{c}\~{a}o de Amparo \`{a} Pesquisa do Estado de S\~{a}o
Paulo (FAPESP), Coordena\c{c}\~{a}o de Aperfei\c{c}oamento do Pessoal
do Nivel Superior (CAPES: AUX-PE-PROCAD 579/2008) and
CNPq/PRONEX/FAPESQ. The work by A. Yu. P. has been supported by the
CNPq project No. 
303461/2009-8.

\end{document}